\documentclass[prl,twocolumn,preprintnumbers,amsmath,amssymb,  superscriptaddress, nobalancelastpage, byrevtex]{revtex4}
\usepackage{graphicx}
\usepackage{subfigure}
\usepackage{dcolumn}
\newcommand{\wcm}{{Wcm$^{-2}$}}
\begin{document}
\title{Excitonic Mott transition in type-II quantum dots}
\author{{Bhavtosh Bansal}\footnote{Present Address: High Field
Magnet Laboratory, Radboud University, Nijmegen 6525 ED, The
Netherlands.}}\email{bhavtosh.bansal@science.ru.nl}
\affiliation{INPAC-Institute for Nanoscale Physics and Chemistry,
Pulsed Fields Group, Katholieke Universiteit Leuven,
Celestijnenlaan 200D, Leuven B-3001, Belgium}
\author{M. Hayne}
\affiliation{INPAC-Institute for Nanoscale Physics and Chemistry,
Pulsed Fields Group, Katholieke Universiteit Leuven,
Celestijnenlaan 200D, Leuven B-3001, Belgium}
\affiliation{Department of Physics, Lancaster University,
Lancaster LA1 4YB, UK.}
\author{M. Geller}
 \affiliation{Institut f\"ur Festk\"orperphysik, PN 5-2,
Hardenbergstr. 36, 10623 Berlin, Germany}
\author{D. Bimberg}
 \affiliation{Institut f\"ur Festk\"orperphysik, PN 5-2,
Hardenbergstr. 36, 10623 Berlin, Germany}
\author{V. V. Moshchalkov}
\affiliation{INPAC-Institute for Nanoscale Physics and Chemistry,
Pulsed Fields Group, Katholieke Universiteit Leuven,
Celestijnenlaan 200D, Leuven B-3001, Belgium}
\date{\today}
\begin{abstract}
Photoluminescence spectra measured on a type-II GaSb/GaAs quantum
dot ensemble at high excitation power indicate a Mott transition
from the low density state comprising of spatially-indirect
excitons to a high density electron-plasma state. Under the
influence of a very high magnetic field, the electron-plasma that
is formed at high excitation powers is `frozen-out' into a state
of optically inactive magneto-excitons.
\end{abstract}
\pacs{71.35.Lk, 78.67.Hc, 71.35.Ji} \maketitle Photoexcited
carriers in a semiconductor are a model system where we have ready
means to both controllably generate and, through the recombination
luminescence, probe the energy levels and statistics of the
exciton population. At low excitation power, the Wannier excitons
can be treated as independent, much like a gas of non-interacting
hydrogenic atoms. A larger density of such entities can lead to
the formation of molecular, liquid, metallic, or superfluid
phases\cite{Zimmermann many particle theory, butov review}. The
low and high density regimes are also separated by a Mott
transition between the ``insulating" excitonic state of bound
electron-hole pairs and the plasma state, which occurs due to the
combination of enhanced screening, bandgap remormalization, and
non-zero bandwidths at high carrier densities. While the dynamics
of photoexcited carriers, the nature of the non-equilibrium
distribution functions, excitonic dephasing, and the transition
itself continue to be actively studied, mainly through
time-resolved studies, in bulk and low-dimensional semiconductor
samples\cite{manzke, vina, shah, fenton, Kappei}, there has also
been an increasing interest in probing the phases of ``spatially
indirect" excitons in systems like type-II quantum wells and
coupled quantum wells with complimentary doping\cite{butov
review}.

We have studied the photoluminescence (PL) spectra from a type-II
self-assembled quantum dot (QD) ensemble. Due to the staggered
band alignment at the GaSb-GaAs interface\cite{manus apl, Geller
APL, hatami, muller-kirsch}, the GaSb QDs contained within the
GaAs matrix\cite{muller-kirsch} confine the holes\cite{Geller
APL}, but act as antidots for the electrons [Fig.1]. Under
non-resonant optical excitation, the electron-hole pairs generated
are spatially separated, with the electrons being loosely bound to
the charge of the holes within the QDs\cite{manus apl, Geller APL,
hatami}. As a result, we have yet another kind of
spatially-indirect exciton where the charge distribution of the
excluded electrons is self-consistently determined such that it
corresponds to the minimum energy configuration\cite{janssens
peeters}, possibly with interesting topological
consequences\cite{ribeiro, govorov, janssens peeters}. These
excitons have no centre of mass degree of freedom and the average
charge of the QDs can be tuned by the strength of the optical
excitation\cite{manus apl, biexciton apl, manus prb}. They do not
form excitonic ``molecules"; the biexcitons are rather like new
``elements" with different ``nuclear" charge\cite{biexciton apl,
manus apl}. The upper bound on the exciton density is determined
by the spatial density of the randomly positioned QDs. Studying
the PL spectra from GaSb QDs at different excitation powers and in
high magnetic fields, we have encountered two Mott transitions.
These are discussed here.

\begin{figure}[b]
\begin{center}
\includegraphics[width=7cm,
  keepaspectratio,
  angle=0,
  origin=lB]{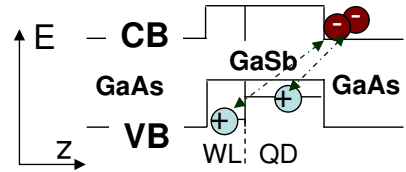}
\caption{\label{fig:simulation}{Simplified band diagram showing
spatially indirect excitons with electrons in the GaAs matrix and
holes in the GaSb valence band of the WL and QD. z-axis denotes
the growth direction and the QDs are distributed in the xy-plane.
Note that the electrons of the WL excitons may screen the Coulomb
interaction of the QD exciton and vice-versa. The band diagram
from an 8-band {\bf k}$\cdot${\bf p} calculation can be found in
reference \onlinecite{manus apl}. The strength of the excitonic
binding is weakened by the presence of the barrier potential.}}
\end{center}
\end{figure}

\begin{figure}[b]
\begin{center}
\includegraphics[width=7.8cm,
  keepaspectratio,
  angle=0,
  origin=lB]{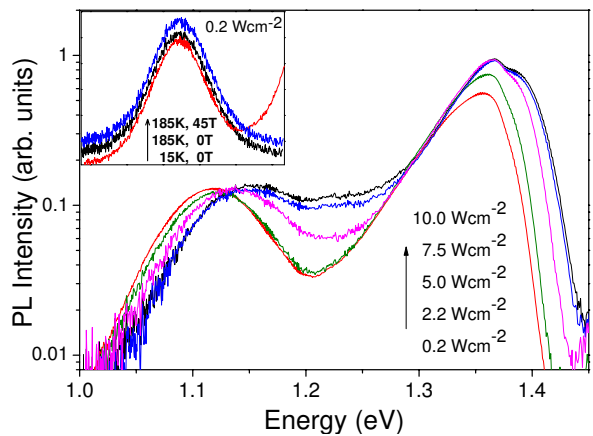}
\caption{\label{fig:simulation}{The zero-field PL spectra at
different excitation powers measured at 15K. The emission around
1.1 eV is from QDs and the higher energy peak is due to the WL.
The QD spectra considerably broaden at higher excitation powers,
indicating the loss of excitonic resonance. Relative to the QD
emission, the WL emission gets stronger as the excitation power is
increased. The asymmetry and shoulder in the WL peak is due to the
sharp cut-off in the detector response at $\sim$ 1.38 eV. (inset)A
comparison of 0.2 \wcm $\,$ spectra (scaled and offset both on the
energy and the intensity axes and now plotted on linear scale),
measured at 15 K (B=0 T) and 185 K (B=0 T, 45 T), show that at
this low power the emission linewidth and the spectral shape are
completely determined by the inhomogeneous broadening and are
independent of magnetic field and temperature. The lineshape is
Gaussian. The WL emission disappears above around 60 K. }}
\end{center}
\end{figure}
The details of sample growth by metal-organic vapor phase epitaxy
and their characterization are described in reference
\onlinecite{muller-kirsch}. The particular sample measured in this
study corresponds to a growth interruption of 2 seconds and a QD
density of about $3\times 10^{10}\, \textrm{cm}^{-2}$
\cite{muller-kirsch}. The low temperature PL spectra exhibit two
prominent peaks, from the wetting layer (WL) at about 900 nm and
the QD ensemble at about 1100 nm [Fig.2]. The ratio of the WL to
the QD PL intensities is strongly excitation power-dependent. This
gives an indication of the relative populations of carriers in the
two regions. In the range of excitation powers studied
here\cite{footnote oidd}, there was a monotonic blue-shift of the
PL peak position with increasing excitation power [Fig. 3(a)].
Unfortunately, this contribution of the capacitive charging energy
associated with the increased hole occupancy at higher excitation
powers\cite{manus apl, biexciton apl} completely masks the bandgap
renormalization effects.

This letter describes the results of non-resonant PL measurements
(excitation wavelength of 514 nm) in high magnetic fields of
$\leq$45 T. All the magnetic field-dependent measurements
described in this study were performed in pulsed magnetic fields
and using a liquid nitrogen-cooled InGaAs linear array detector.
The detector's exposure time was chosen depending on the emission
intensity. The number of exposures per field pulse were determined
by the condition that the variation of the magnetic field B,
$[\sqrt{\langle B^2\rangle-\langle B\rangle^2}/\langle B\rangle]$
during the exposure was less than 5\%. Spectra were measured in
both the rising and the trailing edges and it was ensured that the
magnetic field-dependent change in the peak-position of the PL
spectra (henceforth called PL peak-shift) inferred from spectra
taken in different field pulses had considerable overlaps in the
magnetic field ranges they covered.

The PL peak-shift contains valuable information about the
excitonic states. The magnetic field-induced shift of excitonic
(hydrogenic) ground state at very low fields corresponds to a
diamagnetic shift that is to the lowest order in perturbation
theory quadratic in the magnetic field, $\Delta E\sim
[{q^2\langle\rho^2\rangle/ 8 \mu}]B^2$. The high field regime is
equivalent to having ``free" electrons and holes recombining from
their respective Landau level ground states. Thus the shift is
linear at very high fields, $\Delta E\sim [\hbar q/ 2\mu]B$. q is
absolute value of the electron charge, $\mu$ the reduced mass of
the electron-hole pair, and $\sqrt{\langle\rho^2\rangle}$ is the
exciton radius. These functional forms are strictly valid only in
the two asymptotic limits, but are a good
approximation\cite{Landau_Level_Spectroscopy} for $\beta \lesssim
0.4$ and $\beta \gtrsim 1$ respectively, if $\beta$ is the ratio
of the cyclotron energy to twice the effective exciton Rydberg
energy. As the simplest approximation, a functional form for the
PL peak-shift for $0.4\lesssim\beta\lesssim1.0$ may be
extrapolated by demanding adiabatic continuity between the two
regimes and assuming that diamagnetic contribution extends up to
$\beta=1$. This leads to the following phenomenologically
successful relationship [e.g. Ref. \onlinecite{Schildermans}] to
describe the PL peak-shift for all $\beta$:
\begin{eqnarray}\label{eqn: FIPS}
\Delta E={q^2\langle\rho^2\rangle\over 8 \mu}B^2,\,\,\,
\textrm{for} B<B_c \nonumber
\;\;\;\;\;\;\;\;\;\;\;\;\;\;\;(1\textrm{a})\\
 \Delta E=-{\hbar^2\over
2\mu\langle\rho^2\rangle}+{\hbar q\over 2\mu }B, \,\,\,
\textrm{for} B>B_c \nonumber \,\,\,\,\,\;\;\;(1\textrm{b})
\end{eqnarray}
The crossover field $B_c=2\hbar/ (q\langle\rho^2\rangle)$. The
first term in Eq.(\ref{eqn: FIPS}b) corresponds to the excitonic
binding energy and is the extrapolation of the high-field slope
[second term in Eq.(\ref{eqn: FIPS}b)] to $B=0$.

Fig.3(a) shows the PL peak-shift at different powers of the
excitation laser measured in magnetic fields of up to about $15$ T
with the sample at 4.2 K. Fitting these data to Eq.\ref{eqn: FIPS}
gives a direct way to measure the excitonic binding energies and
the radii. The results are shown in Fig.3(c). The value of the
average binding energy of the exciton decreases from 4.5 meV to
less than 0.5 meV as the excitation power is increased by one and
half orders of magnitude. The evolution is accompanied by the
corresponding increase in the exciton radius from 8 nm to about 25
nm. We thus have an indication of the system progressing toward a
Mott transition\cite{footnote_Fock-Darwin}. The peak-shift
observations correlate well with the zero-field emission spectra
of Fig.2, where despite the large inhomogeneous broadening, the
loss of excitonic resonance is clearly observed by the broadening
of the high energy tail and the significant increase in linewidth
at high excitation power [also see Fig.4(d)]. Reference
\onlinecite{muller-kirsch_many particle} also reports a similar
broadening.

\begin{figure}[b]
\begin{center}
\includegraphics[width=8.7cm,
  keepaspectratio,
  angle=0,
  origin=lB]{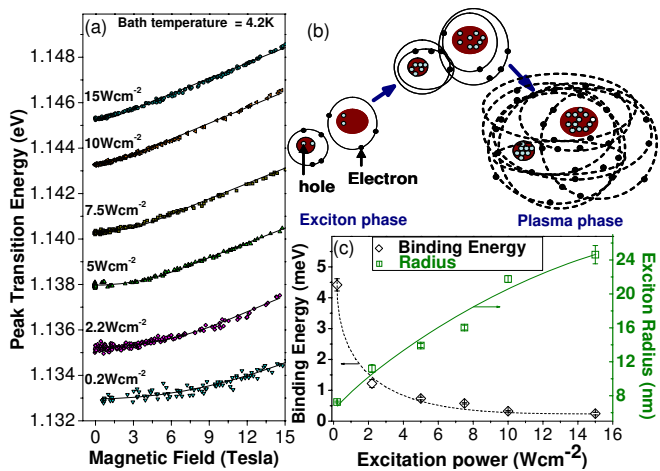}
\caption{\label{fig:simulation}{(a) PL peak transition energy as a
function of magnetic field at different laser powers. Solid lines
are the fits to the equation $E_0+\Delta E$, where $\Delta E$ is
given by Eq.\ref{eqn: FIPS}. Note that $E_0$, the zero-field
transition energy monotonically increases with excitation power.
(b) A pictorial depiction of the corresponding electron
trajectories (represented by solid lines for bound states and
dotted lines for free particles) as the sample goes through a
density-dependent Mott transition. (c) The values of the exciton
radius (squares) and the exciton binding energy (diamonds) as a
function of excitation power, obtained from the fits in (a). The
smooth curves are just to guide the eye. The binding energy and
the radius at 0.2 \wcm excitation are obtained from fits to the
data in Fig.4(c) since a field of $15$ T is not enough to measure
exciton radii of less than about $10$ nm.}}
\end{center}
\end{figure}
The weakening of the binding of electrons from a specific site can
be related to two effects. Of primary importance is the higher
electron density, especially of those electrons which participate
in the WL PL. Fig.2 shows that the WL PL intensity increases much
faster than the QD PL intensity, implying a greater role of
screening at higher excitation power. Specifically, there exists a
critical screening length $\lambda_c$ beyond which the positively
charged dots do not support bound (exciton) states. While for a 3D
system with Yukawa interaction\cite{Rogers} $\lambda_c$ is $0.84
a_B$,  in the actual QD ensemble the barrier potentials at the
QD-matrix heterointerface further aid ionization. QDs are also not
a homogeneous three-dimensional system because the electrons are
constrained to the upper half of the semi-infinite matrix
(excluded by the WL), but still attracted to the holes in the dots
and the WL. Secondly, since the average separation of the {\it
randomly located} QDs in the present sample is about $50$--$60$
nm, for exciton sizes of about 25 nm there may be an overlap
between some excitons with their neighbors. The delocalization of
a few electrons can further aid in screening, leading to further
ionization.

Fig.4 extends the study to much higher magnetic fields (B$<$45 T).
For practical reasons pertaining to helium turbulence during the
pulse, the high field experiments had to be performed in a helium
flow cryostat at a temperature of about $15$ K. It is likely that
the slight increase in temperature has further aided ionization.
The observation of a very large linewidth in Fig.2, at the
excitation power of 10 \wcm, was indicative of the formation of
the plasma phase. Now notice the anomaly in the PL peak-shift of
the spectra measured at this excitation power [Fig.4(a)]. The PL
peak-shift shows an unexpected change in slope around $15$ T and
it can no longer be described by Eq.1 in the entire field range.
Qualitatively, a linear peak-shift at low fields (indicative of
the electron plasma state with near-zero binding energy [cf. Eq.1]
) transforms to a curve described by Eq.1 (dotted line in
Fig.4(a)). This transition from linear to quadratic
dependence\cite{hidden symmetry footnote} is due to the magnetic
field-induced stabilization\cite{fenton} of the insulating
magnetoexciton state\cite{ribeiro, govorov, janssens peeters}---an
arbitrarily weak 1D (effective dimensionality is reduced on
account of 2D magnetic confinement) potential must support at
least one bound state. This is similar to the field-induced
impurity freeze-out phenomenon\cite{fenton} of semiconductor
physics.
 Between Fig.4(a) and (b)
(and other measurements at intermediate powers) we observe an
increase in slope of the field dependence of peak-shift at low
magnetic fields. This suggests a coexistence of two phases, due to
distribution of critical screening lengths for different sized-QD
in the ensemble. Note that Fig.4(c), where we began with a clearly
bound excitonic state at B=0, Eq.1 [solid line in Fig.4(c)]
describes the PL peak-shift in the entire field range and there is
no Mott transition.

Independent evidence of the above described phenomena is provided
by the difference in the field dependence of the linewidths of the
spectra measured at different excitation powers [Fig.2(inset) and
Fig.4(d)]. The linewidth is nearly constant up to $45$ T for the
spectra measured at the excitation power of 0.2 and 1 \wcm,
completely determined by the inhomogeneous broadening.
Fig.2(inset) shows that increasing the measurement temperature to
as high as $185$ K also does not affect the spectral shape.

On the other hand, there is a $\sim 30\%$ decrease in the
linewidth of the spectra measured at 10 \wcm [Fig.4(d)] in an
applied magnetic field. The linewidth is almost constant in plasma
phase below $15$ T and the decrease correlates with the anomaly in
the peak-shift data. Furthermore, in high magnetic field, the
linewidth again tends toward the inhomogeneous broadening-limited
value measured for the excitons at lower laser powers. Note that
this behavior is hard to explain if we assume that the linewidth
enhancement is due to state filling alone\cite{muller-kirsch_many
particle}.

Fig.4(d) thus serves the purpose of a rough ``phase diagram",
where the upper part of the graph corresponding to large
linewidths denotes the plasma phase and the lower part the
(inhomogeneously broadened) exciton phase. It is clarified that
the Mott transition is a zero-temperature phenomenon and thus the
measurements described in Fig.4(d) and the rest of the paper only
depict magnetic field-, excitation power- and temperature-aided
excursions to various phases.

The highly non-equilibrium nature of the phenomena requires the
generation of a large number of photoexcited carriers and
consequently sample excitation under high laser powers, and will
result in some sample heating. However, this heating is only a
perturbation to the experimental conditions: Firstly, the
zero-magnetic field PL spectra showed a strong temperature
dependence at all laser powers and the zero-field PL spectra
measured at different powers all merged to a single curve beyond
100 K (not shown). These observations indicate that the difference
in the sample heating was of the order of about 10K between the
lowest and the highest powers. Furthermore, we observe in Fig.2
that the the loss of excitonic resonance due to many body effects
which change the functional form of the spectral density is
different from the thermal ionization of excitons. For the spectra
measured at 185K and the excitation power of $0.2$ \wcm, the
inhomogeneous linewidth is unaffected despite the large
temperature change.

Finally, we mention that the emission intensities were magnetic
field-dependent. Unlike for type-I dots, we observed a magnetic
field-induced PL quenching, which suggests that the ground state
of magnetoexcitons in our sample is dark with a possibility of a
gap that can be overcome at high temperatures.

\begin{figure}[t]
\begin{center}
\includegraphics[width=7.7cm,
  keepaspectratio,
  angle=0,
  origin=lB]{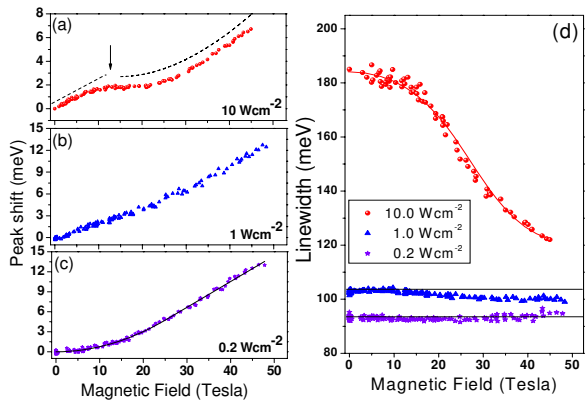}
\caption{\label{fig:simulation}{ PL peak-shift (B$\leq 45$ T) at
different excitation powers. Cryostat temperature $\sim 15$ K. At
high powers, (a) and (b), the slope of the PL peak-shift shows an
unexpected change at intermediate fields. At low power (c), the
conventional excitonic behavior is observed and the PL peak-shift
is well-described by Eq.\ref{eqn: FIPS} (solid line).  (d)
Magnetic field-dependent PL linewidths at different excitation
powers. Solid lines are to guide the eye.}}
\end{center}
\end{figure}

\noindent {\it Conclusions.---}We have studied the different
phases of spatially indirect excitons in an ensemble of GaSb QDs
contained within a GaAs matrix. We have observed systematic
changes in the excitonic binding energy and the exciton Bohr radii
with changing electron-hole density [Fig.3]. These are interpreted
as the transition from an insulating excitonic phase to a single
component plasma phase. The magnetic field induced shift of the
peak energy of the PL spectra, when the system is in the plasma
phase, shows an anomalous functional form [Fig.4] that is
indicative of a phase transition. This is attributed to a second
Mott transition back to the insulating magnetoexciton state. The
loss and the recovery of the excitonic resonance was independently
tracked [Fig.2 and Fig.4(d)] by observing the changes in the
inhomogeneous linewidth as a function of excitation power and
magnetic field.

This work is supported by the Belgian IAP, the SANDiE Network of
Excellence (Contract no. NMP4-CT-2004-500101), and EuroMagNET
(RII3-CT-2004-506239) projects of the European Commission. BB was
a postdoctoral fellow of the FWO Vlaanderen. MH is currently an
Academic Fellow of the Research Councils UK. We thank J. Vanacken
and F. Herlach for their support
and guidance in the pulsed-field experiments.


\begin{references}
\bibitem{footnote oidd}Measurements at ultra-low excitation powers reveal an unexpected U-shape\cite{manus prb} of
the PL peak energy as a function of excitation power, due to
optically induced charge depletion (OICD) of the heavily doped
QDs. The present study is limited to higher powers where we
observe the conventional blue-shift, the physics of the OICD is
not relevant.
\bibitem{footnote_Fock-Darwin}The generic conclusions are model independent. For example, using the Fock-Darwin
Hamiltonian\cite{ribeiro}, instead of Eq.(1), would yield very
similar results.
\end{references}
\end{document}